# Comment on 'Comparative evaluation of two dose optimization methods for image-guided, highly-conformal, tandem and ovoids cervix brachytherapy planning'

Bram L Gorissen[1] and Aswin L Hoffmann[2]

[1]Department of Econometrics and Operations Research/Center for Economic Research (CentER) Tilburg University, PO Box 90153, 5000 LE Tilburg, the Netherlands
[2]Department of Radiation Oncology (MAASTRO), GROW School for Oncology and Developmental Biology, Maastricht University Medical Center, 6201 BN Maastricht, the Netherlands.

With great interest we read the recently published article by Ren *et al*. (2013) on anatomy-based inverse dose optimization for tandem and ovoids cervix brachytherapy planning. The article compares Nelder-Mead Simplex (NMS) with Simulated Annealing (SA) for dwell time optimization, and shows that SA is superior because of better organ-at-risk sparing, lower dwell time variability and smaller sensitivity on the starting point of initial dwell times.

The authors do not distinguish between the optimization algorithm and the optimization model. Both ingredients are needed for inverse treatment planning of the dwell time distribution in HDR-brachytherapy. An optimization model is usually implemented by an objective function $f$ that indicates the quality of a dwell time vector $t$ with a penalty value $f(t)$, where lower penalty values correspond to better dwell time vectors. An optimization algorithm is used to find a solution $t$ such that $f(t)$ is (close to) optimal. Hence, comparing algorithms is only possible based on function values $f(t)$.

Holm et al. (2013) and Gorissen et al. (2013) independently showed that for most commonly used optimization models the penalty value does not correlate well with the plan quality. This becomes more likely when $f$ does not directly incorporate all the evaluation criteria or when the importance factors are not appropriately tuned.

In Ren *et al*. (2013) only the dose distributions are compared, and not the objective function values. It is therefore not clear whether the reported differences in organ-at-risk sparing and dwell time variability were caused by the algorithm or by the model. A graph showing the objective value as a function of the execution time is the only way to objectively compare the algorithms. The conclusion that one algorithm is better than the other cannot be deduced from the results presented by the authors.

Furthermore, the article lacks detailed information about the specific SA implementation. SA is an iterative algorithm where in each step a new dwell time vector is constructed. Without a proper description of the construction procedure and the measures to avoid negative dwell times, it is not possible to reproduce the results presented by the authors.

We invite the authors to respond to both issues.

## References

Gorissen B L, Hertog, D den and Hoffmann A L 2013 Mixed integer programming improves comprehensibility and plan quality in inverse optimization of prostate HDR brachytherapy *Phys. Med. Biol.* 58 1041-58.

Holm Å, Larsson T, Carlsson Tedgren Å 2013 Study of the relationship between dosimetric indices and linear penalties in dose distribution optimization for HDR prostate brachytherapy, in Holm Å 2013, Mathematical optimization of HDR brachytherapy, doctoral dissertation Linköping University, ISSN 0345-7524.

Ren J, Menon G and Sloboda R 2013 Comparative evaluation of two dose optimization methods for image-guided, highly-conformal, tandem and ovoids cervix brachytherapy planning *Phys. Med. Biol.* 58 2045-58.